\documentclass{article}

\usepackage{microtype}
\usepackage{graphicx}
\usepackage{subfigure}
\usepackage{booktabs} 

\usepackage{hyperref}

\usepackage[accepted]{icml2024}

\usepackage{amsmath}
\usepackage{amssymb}
\usepackage{mathtools}
\usepackage{amsthm}

\usepackage[capitalize,noabbrev]{cleveref}

\theoremstyle{plain}

\theoremstyle{definition}

\theoremstyle{remark}

\usepackage[textsize=tiny]{todonotes}

\icmltitlerunning{Physics Constrained Deep Learning For Turbulence}

\begin{document}

\twocolumn[
\icmltitle{Physics Constrained Deep Learning For \\ Turbulence Model Uncertainty Quantification}



\icmlsetsymbol{equal}{*}

\begin{icmlauthorlist}
\icmlauthor{Minghan Chu}{equal,yyy}
\icmlauthor{Weicheng Qian}{equal,comp}
\end{icmlauthorlist}

\icmlaffiliation{yyy}{Department of Mechanical and Materials Engineering, Queen's University, Canada}
\icmlaffiliation{comp}{Department of Computer Science, University of Saskatchewwan, Canada}

\icmlcorrespondingauthor{Minghan Chu}{17MC93@queensu.ca}
\icmlcorrespondingauthor{Weicheng Qian}{weicheng.qian@usask.ca}

\icmlkeywords{Machine Learning, Physics Constrained Deep Learning, Computational Fluid Dynamics, Turbulence Modeling, Uncertainty Quantification, Aerospace Engineering}]



\printAffiliationsAndNotice{}  

\begin{abstract}
Engineering design and scientific analysis rely upon computer simulations of turbulent fluid flows using turbulence models. These turbulence models are empirical and approximate, leading to large uncertainties in their predictions that hamper scientific and engineering advances. We outline a Physics Constrained Deep Learning framework to estimate turbulence model uncertainties using physics based Eigenspace Perturbations along with Deep Learning based guidance. The Deep Learning based modulation controls the spatial variation in perturbation magnitude to improve the calibration of uncertainty estimates over the state of the art physics based methods.
\end{abstract}

\section{Background And Introduction}
\label{Sec1}
Turbulence is a regime of fluid flow exhibiting large stochastic oscillations in velocity and pressure, nonlinear coupling between different scales of fluid motion, increase in mixing, etc \cite{pope2001turbulent}. Turbulence is a central problem across different fields of science and engineering including Aerospace Engineering, Astrophysics, Bio-medicine, Environmental Engineering, etc \cite{tennekes1972first}. The ability to reliably and accurately predict the behavior and evolution of fluid turbulence will lead to advances in all these disciplines. For example, better predictions of the turbulence in the cardiovascular system can help us design improved artificial valves for patients. Despite years of research there is no analytical theory to predict the evolution of turbulent flows. In engineering design and scientific analysis where turbulent flows are common researchers have to use turbulence models for turbulent flow simulations on computers. Turbulence models are simplified empirical constitutive relations attempting to capture some features of turbulence physics \cite{speziale1991modelling, launder1975progress}. The approximations of turbulence models lead to large uncertainties in their predictions \cite{smith2013uncertainty}. The uncertainties have deleterious effects such as engineering designs that are not efficient or even safe for deployment. Estimating turbulence model uncertainties is described as the most critical problem in aerospace design \cite{kato2013approach}. 

The only method to estimate turbulence model uncertainties currently is the Eigenspace Perturbation Method \cite{iaccarino2017eigenspace} using injected perturbations in the model's predictions to estimate the variability in quantities of interest like thrust, lift, drag, etc. This method has been applied in a range of engineering problems like virtual certification of aircraft designs \cite{mukhopadhaya2020multi, nigam2021toolset}, design of civil structures \cite{gorle2019epistemic, thompson2019eigenvector}, aerospace design\cite{mishra2019uncertainty, mishra2017rans, mishra2019estimating, mishra2017uncertainty}, design under uncertainty (DUU) and reliability based design optimization\cite{demir2023robust, cook2019optimization, mishra2020design, righi2023uncertainties}, etc. Being purely physics based, the EMP assigns equal weight to all possible turbulence states, ignoring their relative likelihood for the flow under consideration. Similarly, the EMP uses the worst case perturbation over the entire fluid flow domain, instead to concentrating the perturbations in regions where turbulence modeling simplifications are imprecise like regions of flow separation. These leads to overly conservative and ill calibrated uncertainty bounds. A alleviate this, a marker function is required that can modulate the perturbation magnitude for different flows and for different regions of the same flow. Such a function while not analytically or physically feasible can be learned from data. Machine Learning based approaches are becoming more used in fluid mechanics applications \cite{duraisamy2019turbulence, chung2021data, chung2022interpretable, brunton2020machine}. Prior investigations have used ML to improve turbulence model UQ \cite{xiao2016quantifying,wu2018physics,heyse2021estimating,heyse2021data,zeng2022adaptive}. In our Physics Constrained Deep Learning approach, we attempt to use the foundations of the EPM and augment it using domain knowledge informed ML models. Here we identify the absence of non-local modeling information as a key delimiter and utilize Convolutional Neural Network models for modeling the marker function from data.

\section{Methodology}

\begin{figure*}[t]
    \centering
    \includegraphics[width=0.7\linewidth]{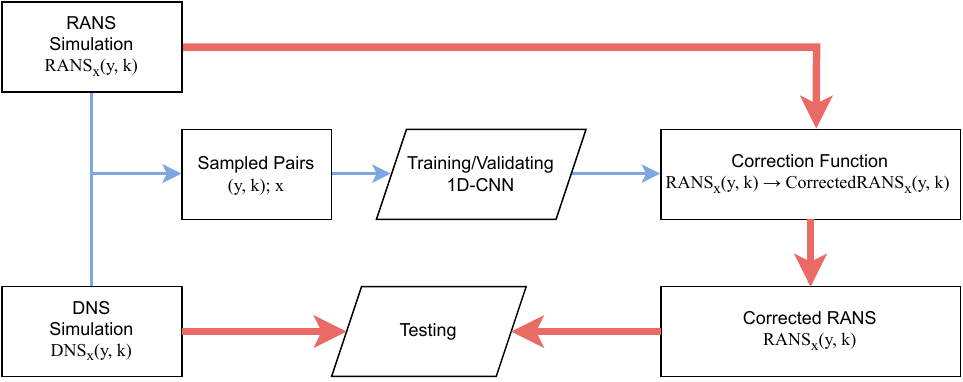}
    \caption{The data flow and methodology used in this investigation. The blue path is the training path,the red path is the validation path.} 
    \label{fig:data-flow.pdf}
\end{figure*}

Turbulent flows are characterized by stochastic fluctuations in the velocity field. The covariance of these fluctuations is the Reynolds stress tensor, $R_{ij}$. The turbulence kinetic energy is the trace of this tensor and is a good indicator of the accuracy of the turbulence model\cite{iaccarino2017eigenspace}. We use a one-dimensional convolutional neural network (1D-CNN) to learn the function to correct the lower fidelity turbulence model prediction (hereafter, RANS for Reynolds Averaged Navier Stokes model) to the high fidelity data (hereafter, DNS for Direct Numerical Simulation). 

For both the RANS and DNS simulations we summarize their results as the function of the perturbed turbulence kinetic energy $k^{*} = f(x,y)$, where $x$ and $y$ are coordinates in a two-dimensional computational domain, and $f$ is the mapping from every coordinate $(x, y)$ to $k^{*}$, expressed as tuples $(x, y, k^{*})$ from simulation results. The correction function for RANS is a mapping between two functions:
\begin{equation}
Z: f^{\mathrm{RANS}}(x,y) \rightarrow f^{\mathrm{DNS}}(x,y)
\end{equation}
with $k^{\mathrm{DNS}} = f^{\mathrm{DNS}}(x, y)$ and $k^{\mathrm{RANS}} = f^{\mathrm{RANS}}(x, y)$, we can rewrite $Z$ as a mapping $\zeta$ between points that comprises $f^{\mathrm{RANS}}$ and $f^{\mathrm{DNS}}$

\begin{equation}
\zeta: (x, y, k^{\mathrm{RANS}}) \rightarrow (x, y, k^{\mathrm{DNS}})
\end{equation}

Considering the model error for RANS and DNS in terms of turbulence kinetic energy, we have

\begin{equation}
   p^{\text {RANS }}\left(K_g \mid x, y\right)=p\left(k_g=k^{\text {RANS }} \mid x, y\right)
\end{equation}

\begin{equation}
    p^{\text {DNS }}\left(K_g \mid x, y\right)=p\left(k_g=k^{\text {DNS }} \mid x, y\right)
\end{equation}

where $K_g$ is the unknown ground truth of turbulence kinetic energy at $(x, y)$. Turbulence kinetic energy resulted from DNS simulation results $p^{\mathrm{RANS}}$ can be estimated with turbulence kinetic energy from RANS simulation $p^{\mathrm{DNS}}$ and its correction function $g$ as

\begin{equation}
p^{\mathrm{DNS}}\left(K_g \mid x, y\right)=g\left(k^{\mathrm{RANS}}, x, y\right) p\left(k^{\mathrm{RANS}} \mid x, y\right)
\end{equation}

As $k^{\mathrm{DNS}} = f^{\mathrm{DNS}}(x, y)$ and $k^{\mathrm{RANS}} = f^{\mathrm{RANS}}(x, y)$, at each $x$, we have that $k_x^{\mathrm{DNS}} = f_x^{\mathrm{DNS}}(y)$ and $k_x^{\mathrm{RANS}} = f_x^{\mathrm{RANS}}(y)$, assuming both $f_x^{\mathrm{RANS}}$ and $f_x^{\mathrm{DNS}}$ are continuous. In other words, we can learn $\hat{g}$ with paired $(\mathbf{k}_{x,y,\delta}^{\mathrm{RANS}}, \mathbf{k}_{x,y,\delta}^{\mathrm{DNS}})$.

\section{Experiments Setup and Data Sources}
Our objective is to learn from data a mapping from the space of lower fidelity turbulence model simulations (RANS) to high fidelity ground truth data (DNS). To learn this correction function, we utilize a CNN to develop this ``correction function" on two diverse paired datasets: a flow over an airfoil at a high angle of attack \cite{zhang2021turbulent,chu2022model} and the public dataset of a channel flow over periodic hills \cite{voet2021hybrid}. Both these flow cases experience adverse pressure gradient, which causes the complex flow features of separation and reattachment. Such complex behaviors have been the bane of turbulence models and make these datasets relevant to our objectives. 

In \cref{fig:data-flow.pdf}, we show our split of $x$-coordinate grouped pairs of $(\mathbf{k}_{x,y,\delta}^{\mathrm{RANS}}, \mathbf{k}_{x,y,\delta}^{\mathrm{DNS}})$ into a training set and a validating set by their group key $x$. For both the datasets, we choose $x$ at only three positions on the geometry from the beginning, the middle, and the end of all paired $x$ values. For the in-house dataset based on the flow over an airfoil, $x/c = 0.4, 0.56, 0.58$; for the public dataset based on the two-dimensional flow over periodic hills, $x/c = 0, 0.046, 0.116, 0.128$, where $c$ is a reference length used for normalization. For each dataset, we use a 80\%--20\% split as training-testing dataset. 

To train the model, we use the Mean Absolute Error (MAE) as the objective function. In contrast to the Mean Squared Error (MSE or $L_2$ loss) the MAE (or the $L_1$ loss) does not penalize incorrect predictions as heavily and better final models were developed in this study using the MAE loss as the objective function. The MAE loss is computed for the uncorrected RANS and the MAE loss of 1D-CNN corrected RANS. These are compared to exhibit the efficacy of our approach.

\section{Results}
\begin{figure*}[!htb]
    \centering
    \includegraphics[width=\linewidth]{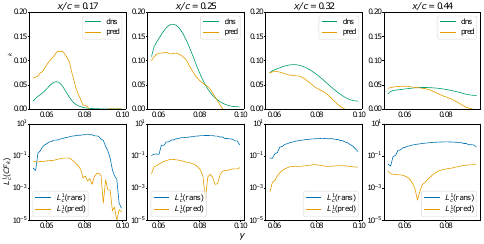}
    \caption{Results for Selig-Donovan 7003 airfoil. First row: CNN corrected DNS (\texttt{pred}) compared with ground truth (\texttt{dns}). Second row: Validation of 1D-CNN by comparing L1 loss between $L^1_c(\texttt{rans})$ and $L^1_c(\texttt{pred})$.}
    \label{fig:cnn-corrected-rans-zhang.pdf}
\end{figure*}

\begin{figure*}[!htb]
    \centering
    \includegraphics[width=\linewidth]{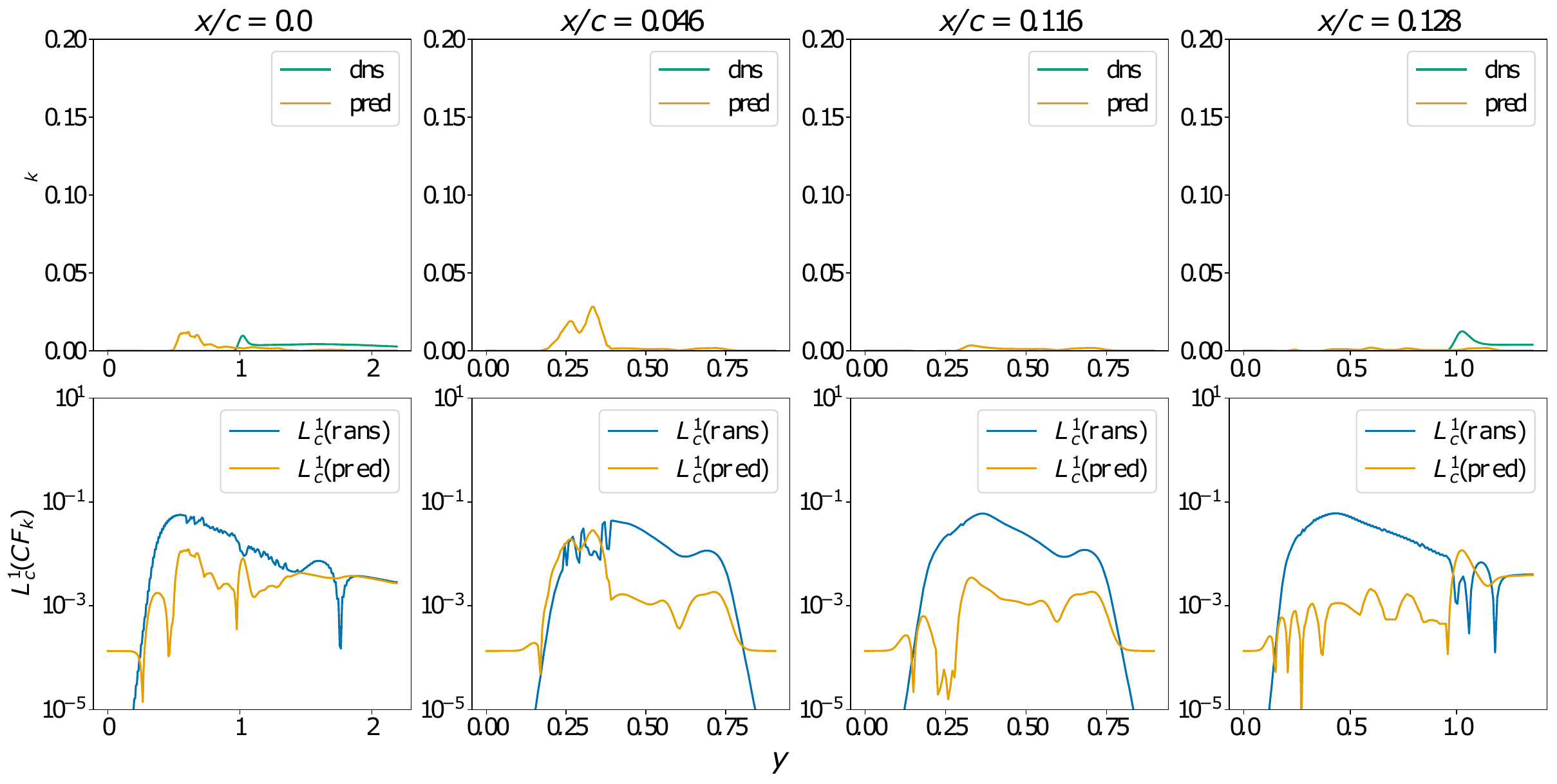}
    \caption{Results for two-dimensional periodic hills. First row: CNN-based prediction (\texttt{pred}) compared with ground truth (\texttt{dns}). Second row: Validation of 1D-CNN by comparing L1 loss between $L^1_c(\texttt{rans})$ and $L^1_c(\texttt{pred})$.}
    \label{fig:cnn-corrected-rans-voet.pdf}
\end{figure*}

The CNN model's performance is validated at all paired $x$ locations. \Cref{fig:cnn-corrected-rans-zhang.pdf,fig:cnn-corrected-rans-voet.pdf} show the results for the flow over an airfoil and over 2D periodic hills, respectively, at four $x$ locations within the region of separated flow. From \cref{fig:cnn-corrected-rans-zhang.pdf}, the series of CNN predicted DNS profiles in the first row are then smoothed with the moving average with a window size of six consecutive estimations. The CNN model prediction for the turbulence kinetic energy profile approximates the ground truth DNS despite being trained on a small dataset. 

For both datasets, the CNN correction function is trained on paired RANS-DNS simulated turbulence kinetic energy using less than $20\%$ positions along $x$-axis, while the CNN function is still effective for the remaining $80\%$ positions. Effectively, the CNN correction can be used to predict RANS-DNS simulated turbulence kinetic energy at any $x$ coordinate with only a fraction of the whole $x$ coordinates. Furthermore, our lightweight CNN model uses the $y$ coordinates for grouping RANS-simulated turbulence kinetic energy within a neighbor. The results of the CNN correction function suggests that RANS results might be improved by leveraging information embedded in the positions within a close neighbor, which is independent of the absolute coordinates $(x, y)$. The CNN correction function trained on one case can still help smooth and reduce the error of RANS results for other cases.

For both the in-house and the public flow cases, our CNN prediction for $k$ lies closer to the DNS data at any $x$ location, i.e., the discrepancy in general reduces as the flow proceeds further downstream. The RANS results deviate from the DNS data in both flow scenarios and our CNN-based correction function can significantly reduce the L1 error of RANS- from DNS-simulations, as shown in the second row of \cref{fig:cnn-corrected-rans-zhang.pdf,fig:cnn-corrected-rans-voet.pdf}, i.e., a $L^1_c(\texttt{pred})$ drop of two orders compared to $L^1_c(\texttt{rans})$. 

From \Cref{fig:cnn-corrected-rans-zhang.pdf,fig:cnn-corrected-rans-voet.pdf}, it is important to note that the CNN-based prediction for $k$ tends to approach closer to the DNS profile as the flow proceeds further downstream. This indicates that the CNN-based correction function tends to become more trustworthy within the region of fully turbulent flow where flow features are less complex than that within region of separated flow where rather complex flow features evolve. From the \cref{fig:cnn-corrected-rans-zhang.pdf}, the second row shows the computed $L_1$ error of the RANS-based prediction and the CNN-based prediction. It is clear that the $L_1$ error for CNN-based correction function can significantly reduce the $L_1$ error in magnitude compared to that for the original RANS. 


\section{Discussion}
We proposed a Convolutional Neural Network based method to approximate the correction function that corrects low fidelity turbulence model simulation predictions towards high fidelity data. We further examined our method on two datasets: 1) one flow being considered is over a SD7003 airfoil at $8^\circ$ angle of attack and the Reynolds number based on the cord length of $Re_{c} = 60000$ \cite{zhang2021turbulent}. A laminar separation bubble evolves on the suction side of the airfoil whereby the flow undergoes transition to turbulence, 2) another is generated from the DNS two-dimensional channel flow over periodic hills \cite{voet2021hybrid}. The flow experiences adverse pressure gradient when encountering the curved surface of the hill. It should be noted that a separation bubble occurs for both datasets. The RANS results deviate from the DNS data in both flow scenarios and our CNN-based correction function can significantly reduce the L1 error of RANS- from DNS-simulations.

1. Our lightweight CNN model uses only the RANS-simulated turbulence kinetic energy at neighbor on the $y$-axis to correct RANS towards DNS simulation results. The results showed great improvement compared to the DNS data. 
2. Our light-weight model trained on the in-house DNS dataset is directly used to yield corrected RANS predictions for the periodically arranged hills, and show a clear improvement. A separated region in both flow cases might partly explain this improvement. 
3. Also similar cross neighbor features might contribute this improvement. 

For both datasets, the CNN-based correction function is trained on paired RANS-DNS simulated turbulence kinetic energy using less than $20\%$ positions along $x$-axis, while the CNN-based function is still effective for the remaining $80\%$ positions. In other words, our CNN-based correction can be used to predict RANS-DNS simulated turbulence kinetic energy at any $x$ coordinate with only a fraction of the whole $x$ coordinates. Furthermore, our lightweight CNN model uses the $y$ coordinates for grouping RANS-simulated turbulence kinetic energy within a neighbor. The results of our CNN-based correction function suggests that RANS results might be improved by leveraging information embedded in the positions within a close neighbor, which is independent of the absolute coordinates $(x, y)$.

There are relatively few studies for correcting the perturbed turbulence kinetic energy. Very recently, the study of Chu \textit{et al.} \cite{chu2022model} assessed the effect of polynomial regression on the estimation of the perturbed turbulence kinetic energy. Our CNN-based correction method has readily implications on practical applications, such as, to be coupled to the eigenspace perturbation approach of Emory \textit{et al.} \cite{emory2013modeling}. The eigenspace perturbation approach has been implemented within the OpenFOAM framework to construct a marker function for the perturbed turbulence kinetic energy \cite{chu2022model}. Our CNN-based correction method can be used as a new marker function to predict the perturbed turbulence kinetic energy. 

\section{Application of the lightweight CNN-based correction function on UQ for an SD 7003 airfoil}
Our CNN-based correction function method can be applied to different flow cases to correct RANS towards DNS. In this section, the CNN-based correction function is applied to the SD7003 airfoil case to predict the perturbed turbulence kinetic energy.

\begin{figure*}[t]
         \centering
         \includegraphics[width=\textwidth]{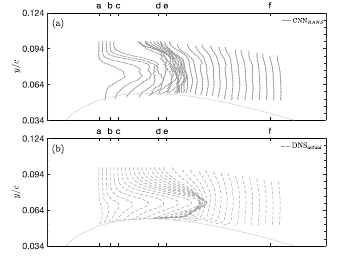}
        \caption{(a) CNN corrected RANS (\texttt{CNN\_{DNS}}) {(solid-dotted lines)} of the normalized perturbed turbulence kinetic energy and (b) ground truth (\texttt{CNN\_{RANS}}) along the suction side of the SD7003 airfoil (geometry depicted by gray line): from left to right are zone $ab$, zone $cd$ and zone $ef$. There are 32 positions on the suction side of the airfoil.}
        \label{fig:CNN_DNS.pdf}
\end{figure*}

The CNN corrected RANS and ground truth profiles for the turbulence kinetic energy normalized with the freestream velocity squared, $k^{*}/U_{\infty}^2$ and $k/U_{\infty}^2$ are shown in Figs. \ref{fig:CNN_DNS.pdf} (a) and (b), respectively. The $k^{*}/U_{\infty}^2$ and $k/U_{\infty}^2$ profiles are equally spaced for the $ab$ and $cd$ zone with $x/c = 0.01$, and a uniform spacing of $x/c = 0.02$ is used for the $ef$ zone. It is clear that the $k^{*}/U_{\infty}^2$ and $k/U_{\infty}^2$ profiles are more densely packed for the $ab$ and $cd$ zone, within which the flow features are complex due to the presence of separation and reattachment. 

From Figs. \ref{fig:CNN_DNS.pdf} (a) and (b), the CNN corrected DNS profiles in general exhibit a similar trend as that for the ground truth dataset, as both profiles show a gradual increase in the $ab$ and $cd$ zone. Then a reduction of the profile is observed further downstream in the $ef$ zone. 

In Fig. \ref{fig:CNN_DNS.pdf} (a), CNN corrected RANS profiles in general increase in magnitude as the flow moves further downstream, which is qualitatively similar to the ground truth profiles. Further, it should be noted that the CNN corrected RANS profiles increase in a somewhat larger magnitude than that for the ground truth in the $ab$ zone. The discrepancy is more than $50\%$ at the beginning of the $ab$ zone and gradually reduces as the flow moves further downstream, which indicates that a better accuracy of our CNN model is yielded further downstream. This behavior becomes more clear for the $cd$ and $ef$ zone. In the region where the end of the $cd$ zone meets the beginning of the $ef$ zone, the ground truth profiles are clustered due to the complex flow feature of the reattachment \cite{chu2022quantification}, as shown in Figs \ref{fig:CNN_DNS.pdf} (b). This clustering behavior is successfully captured by our CNN model, as shown in Fig. \ref{fig:CNN_DNS.pdf} (a). In the $ef$ zone, our CNN model gives overall accurate predictions for the $k^{*}/U_{\infty}^{2}$ profiles, i.e., the CNN corrected RANS profiles and the ground truth profiles are almost identical.



\section{Conclusion}
\label{sec:Conclusion}
Engineering and scientific analysis use computer simulations of turbulent fluid flows using turbulence models. Turbulence models are empirical and approximate, leading to large uncertainties in their predictions that delimit their utility. Ascertaining these uncertainties is a longstanding problem in engineering and science, with a need across aerospace engineering, astrophysics, bio-medicine, etc. This investigation presents a Physics Constrained Deep Learning method to solve this problem, using deep convolutional neural networks to formulate a spatially varying marker function that can control the degree of the eigenvalue perturbation in the EPM. In the recent past there have been investigations that seek to utilize Machine Learning models to control the EPM \cite{heyse2021data, heyse2021estimating, matha2023evaluation}. The present study is the first to analyze the projection from RANS prediction space to the space of DNS data using the CNN models. Our use of convolutions allows the inclusion of non-local physics in the uncertainty estimation. The results show that the CNN models learn the discrepancy between RANS simulations and DNS data, resulting in a surrogate model for this marker function.

\section{Impact statement}
Turbulent flows are commonplace and important in problems across different fields of science including Aerospace Engineering, Astrophysics, Bio-medicine, Environmental Engineering, etc. Reliable and accurate predictions of the behavior and evolution of fluid turbulence would lead to advances in all these disciplines. For instance, better predictions of the turbulence in the cardiovascular system can help us design improved artificial valves for patients. Most engineering and science studies use computer simulations of turbulent flow with turbulence models. These models are simple, empirical, approximate and at the end of the day, limited in the fidelity with which they can replicate turbulence physics. Flow separation, reattachment, and separation bubbles are frequently encountered in real life turbulent flows. However, those flow features are complex to capture using the Reynolds Averaged Navier Stokes (RANS) turbulence models in general and in particular using the eddy viscosity based models that are commonly used in engineering and science. 

Alternatives computation approaches like Large Eddy Simulations (LES) and Direct Numerical Simulations (DNS) give high-fidelity results, but the calculations are too expensive for practical engineering applications. Therefore, knowing the uncertainty associated with the low-cost RANS simulations are of great interest as it builds confidence when using RANS. An effective physics-based eigenspace perturbation method only presupposes the worst case scenario where maximum strength (turbulence kinetic energy) of perturbation is assumed. Thus it gives overly conservative uncertainty estimates. Our physics inspired machine learning based perturbation approach is able to spatially modulate the strength and location of the physics based perturbations and leads to improved estimates of turbulence model errors.

\bibliography{examplepaper}

\providecommand{\noopsort}[1]{}\providecommand{\singleletter}[1]{#1}%
\begin{thebibliography}{34}
\providecommand{\natexlab}[1]{#1}
\providecommand{\url}[1]{\texttt{#1}}
\expandafter\ifx\csname urlstyle\endcsname\relax
  \providecommand{\doi}[1]{doi: #1}\else
  \providecommand{\doi}{doi: \begingroup \urlstyle{rm}\Url}\fi

\bibitem[Brunton et~al.(2020)Brunton, Noack, and Koumoutsakos]{brunton2020machine}
Brunton, S.~L., Noack, B.~R., and Koumoutsakos, P.
\newblock Machine learning for fluid mechanics.
\newblock \emph{Annual Review of Fluid Mechanics}, 52:\penalty0 477--508, 2020.

\bibitem[Chu et~al.(2022{\natexlab{a}})Chu, Wu, and Rival]{chu2022model}
Chu, M., Wu, X., and Rival, D.~E.
\newblock Model-form uncertainty quantification of reynolds-averaged navier--stokes modeling of flows over a sd7003 airfoil.
\newblock \emph{Physics of Fluids}, 34\penalty0 (11):\penalty0 117105, 2022{\natexlab{a}}.

\bibitem[Chu et~al.(2022{\natexlab{b}})Chu, Wu, and Rival]{chu2022quantification}
Chu, M., Wu, X., and Rival, D.~E.
\newblock Quantification of reynolds-averaged-navier--stokes model-form uncertainty in transitional boundary layer and airfoil flows.
\newblock \emph{Physics of Fluids}, 34\penalty0 (10):\penalty0 107101, 2022{\natexlab{b}}.

\bibitem[Chung et~al.(2021)Chung, Mishra, Perakis, and Ihme]{chung2021data}
Chung, W.~T., Mishra, A.~A., Perakis, N., and Ihme, M.
\newblock Data-assisted combustion simulations with dynamic submodel assignment using random forests.
\newblock \emph{Combustion and Flame}, 227:\penalty0 172--185, 2021.

\bibitem[Chung et~al.(2022)Chung, Mishra, and Ihme]{chung2022interpretable}
Chung, W.~T., Mishra, A.~A., and Ihme, M.
\newblock Interpretable data-driven methods for subgrid-scale closure in les for transcritical lox/gch4 combustion.
\newblock \emph{Combustion and Flame}, 239:\penalty0 111758, 2022.

\bibitem[Cook et~al.(2019)Cook, Mishra, Jarrett, Willcox, and Iaccarino]{cook2019optimization}
Cook, L.~W., Mishra, A., Jarrett, J., Willcox, K., and Iaccarino, G.
\newblock Optimization under turbulence model uncertainty for aerospace design.
\newblock \emph{Physics of Fluids}, 31\penalty0 (10):\penalty0 105111, 2019.

\bibitem[Demir et~al.(2023)Demir, Gorguluarslan, and Aradag]{demir2023robust}
Demir, G., Gorguluarslan, R.~M., and Aradag, S.
\newblock Robust shape optimization under model uncertainty of an aircraft wing using proper orthogonal decomposition and inductive design exploration method.
\newblock \emph{Structural and Multidisciplinary Optimization}, 66\penalty0 (4):\penalty0 93, 2023.

\bibitem[Duraisamy et~al.(2019)Duraisamy, Iaccarino, and Xiao]{duraisamy2019turbulence}
Duraisamy, K., Iaccarino, G., and Xiao, H.
\newblock Turbulence modeling in the age of data.
\newblock \emph{Annual Review of Fluid Mechanics}, 51:\penalty0 357--377, 2019.

\bibitem[Emory et~al.(2013)Emory, Larsson, and Iaccarino]{emory2013modeling}
Emory, M., Larsson, J., and Iaccarino, G.
\newblock Modeling of structural uncertainties in reynolds-averaged navier-stokes closures.
\newblock \emph{Physics of Fluids}, 25\penalty0 (11):\penalty0 110822, 2013.

\bibitem[Gorl{\'e} et~al.(2019)Gorl{\'e}, Zeoli, Emory, Larsson, and Iaccarino]{gorle2019epistemic}
Gorl{\'e}, C., Zeoli, S., Emory, M., Larsson, J., and Iaccarino, G.
\newblock Epistemic uncertainty quantification for reynolds-averaged navier-stokes modeling of separated flows over streamlined surfaces.
\newblock \emph{Physics of Fluids}, 31\penalty0 (3):\penalty0 035101, 2019.

\bibitem[Heyse et~al.(2021{\natexlab{a}})Heyse, Mishra, and Iaccarino]{heyse2021data}
Heyse, J.~F., Mishra, A.~A., and Iaccarino, G.
\newblock Data driven physics constrained perturbations for turbulence model uncertainty estimation.
\newblock In \emph{AAAI Spring Symposium: MLPS}, 2021{\natexlab{a}}.

\bibitem[Heyse et~al.(2021{\natexlab{b}})Heyse, Mishra, and Iaccarino]{heyse2021estimating}
Heyse, J.~F., Mishra, A.~A., and Iaccarino, G.
\newblock Estimating rans model uncertainty using machine learning.
\newblock \emph{Journal of the Global Power and Propulsion Society}, 2021\penalty0 (May):\penalty0 1--14, 2021{\natexlab{b}}.

\bibitem[Iaccarino et~al.(2017)Iaccarino, Mishra, and Ghili]{iaccarino2017eigenspace}
Iaccarino, G., Mishra, A.~A., and Ghili, S.
\newblock Eigenspace perturbations for uncertainty estimation of single-point turbulence closures.
\newblock \emph{Physical Review Fluids}, 2\penalty0 (2):\penalty0 024605, 2017.

\bibitem[Kato \& Obayashi(2013)Kato and Obayashi]{kato2013approach}
Kato, H. and Obayashi, S.
\newblock Approach for uncertainty of turbulence modeling based on data assimilation technique.
\newblock \emph{Computers \& Fluids}, 85:\penalty0 2--7, 2013.

\bibitem[Launder et~al.(1975)Launder, Reece, and Rodi]{launder1975progress}
Launder, B.~E., Reece, G.~J., and Rodi, W.
\newblock Progress in the development of a reynolds-stress turbulence closure.
\newblock \emph{Journal of fluid mechanics}, 68\penalty0 (3):\penalty0 537--566, 1975.

\bibitem[Matha et~al.(2023)Matha, Kucharczyk, and Morsbach]{matha2023evaluation}
Matha, M., Kucharczyk, K., and Morsbach, C.
\newblock Evaluation of physics constrained data-driven methods for turbulence model uncertainty quantification.
\newblock \emph{Computers \& Fluids}, 255:\penalty0 105837, 2023.

\bibitem[Mishra \& Iaccarino(2017{\natexlab{a}})Mishra and Iaccarino]{mishra2017rans}
Mishra, A. and Iaccarino, G.
\newblock Rans predictions for high-speed flows using enveloping models.
\newblock \emph{arXiv preprint arXiv:1704.01699}, 2017{\natexlab{a}}.

\bibitem[Mishra \& Iaccarino(2017{\natexlab{b}})Mishra and Iaccarino]{mishra2017uncertainty}
Mishra, A.~A. and Iaccarino, G.
\newblock Uncertainty estimation for reynolds-averaged navier--stokes predictions of high-speed aircraft nozzle jets.
\newblock \emph{AIAA Journal}, 55\penalty0 (11):\penalty0 3999--4004, 2017{\natexlab{b}}.

\bibitem[Mishra et~al.(2019{\natexlab{a}})Mishra, Duraisamy, and Iaccarino]{mishra2019estimating}
Mishra, A.~A., Duraisamy, K., and Iaccarino, G.
\newblock Estimating uncertainty in homogeneous turbulence evolution due to coarse-graining.
\newblock \emph{Physics of Fluids}, 31\penalty0 (2):\penalty0 025106, 2019{\natexlab{a}}.

\bibitem[Mishra et~al.(2019{\natexlab{b}})Mishra, Mukhopadhaya, Iaccarino, and Alonso]{mishra2019uncertainty}
Mishra, A.~A., Mukhopadhaya, J., Iaccarino, G., and Alonso, J.
\newblock Uncertainty estimation module for turbulence model predictions in su2.
\newblock \emph{AIAA Journal}, 57\penalty0 (3):\penalty0 1066--1077, 2019{\natexlab{b}}.

\bibitem[Mishra et~al.(2020)Mishra, Mukhopadhaya, Alonso, and Iaccarino]{mishra2020design}
Mishra, A.~A., Mukhopadhaya, J., Alonso, J., and Iaccarino, G.
\newblock Design exploration and optimization under uncertainty.
\newblock \emph{Physics of Fluids}, 32\penalty0 (8):\penalty0 085106, 2020.

\bibitem[Mukhopadhaya et~al.(2020)Mukhopadhaya, Whitehead, Quindlen, Alonso, and Cary]{mukhopadhaya2020multi}
Mukhopadhaya, J., Whitehead, B.~T., Quindlen, J.~F., Alonso, J.~J., and Cary, A.~W.
\newblock Multi-fidelity modeling of probabilistic aerodynamic databases for use in aerospace engineering.
\newblock \emph{International Journal for Uncertainty Quantification}, 10\penalty0 (5), 2020.

\bibitem[Nigam et~al.(2021)Nigam, Mohseni, Valverde, Voronin, Mukhopadhaya, and Alonso]{nigam2021toolset}
Nigam, N., Mohseni, S., Valverde, J., Voronin, S., Mukhopadhaya, J., and Alonso, J.~J.
\newblock A toolset for creation of multi-fidelity probabilistic aerodynamic databases.
\newblock In \emph{AIAA Scitech 2021 Forum}, pp.\  0466, 2021.

\bibitem[Pope(2001)]{pope2001turbulent}
Pope, S.~B.
\newblock Turbulent flows, 2001.

\bibitem[Righi(2023)]{righi2023uncertainties}
Righi, M.
\newblock Uncertainties quantification in the prediction of the aeroelastic response of the pazy wing tunnel model.
\newblock In \emph{AIAA SCITECH 2023 Forum}, pp.\  0761, 2023.

\bibitem[Smith(2013)]{smith2013uncertainty}
Smith, R.~C.
\newblock \emph{Uncertainty quantification: theory, implementation, and applications}, volume~12.
\newblock Siam, 2013.

\bibitem[Speziale et~al.(1991)Speziale, Sarkar, and Gatski]{speziale1991modelling}
Speziale, C.~G., Sarkar, S., and Gatski, T.~B.
\newblock Modelling the pressure--strain correlation of turbulence: an invariant dynamical systems approach.
\newblock \emph{Journal of fluid mechanics}, 227:\penalty0 245--272, 1991.

\bibitem[Tennekes \& Lumley(1972)Tennekes and Lumley]{tennekes1972first}
Tennekes, H. and Lumley, J.~L.
\newblock \emph{A first course in turbulence}.
\newblock MIT press, 1972.

\bibitem[Thompson et~al.(2019)Thompson, Mishra, Iaccarino, Edeling, and Sampaio]{thompson2019eigenvector}
Thompson, R.~L., Mishra, A.~A., Iaccarino, G., Edeling, W., and Sampaio, L.
\newblock Eigenvector perturbation methodology for uncertainty quantification of turbulence models.
\newblock \emph{Physical Review Fluids}, 4\penalty0 (4):\penalty0 044603, 2019.

\bibitem[Voet et~al.(2021)Voet, Ahlfeld, Gaymann, Laizet, and Montomoli]{voet2021hybrid}
Voet, L.~J., Ahlfeld, R., Gaymann, A., Laizet, S., and Montomoli, F.
\newblock A hybrid approach combining dns and rans simulations to quantify uncertainties in turbulence modelling.
\newblock \emph{Applied Mathematical Modelling}, 89:\penalty0 885--906, 2021.

\bibitem[Wu et~al.(2018)Wu, Xiao, and Paterson]{wu2018physics}
Wu, J.-L., Xiao, H., and Paterson, E.
\newblock Physics-informed machine learning approach for augmenting turbulence models: A comprehensive framework.
\newblock \emph{Physical Review Fluids}, 3\penalty0 (7):\penalty0 074602, 2018.

\bibitem[Xiao et~al.(2016)Xiao, Wu, Wang, Sun, and Roy]{xiao2016quantifying}
Xiao, H., Wu, J.-L., Wang, J.-X., Sun, R., and Roy, C.
\newblock Quantifying and reducing model-form uncertainties in reynolds-averaged navier--stokes simulations: A data-driven, physics-informed bayesian approach.
\newblock \emph{Journal of Computational Physics}, 324:\penalty0 115--136, 2016.

\bibitem[Zeng et~al.(2022)Zeng, Zhang, Li, Zhang, and Yan]{zeng2022adaptive}
Zeng, F., Zhang, W., Li, J., Zhang, T., and Yan, C.
\newblock Adaptive model refinement approach for bayesian uncertainty quantification in turbulence model.
\newblock \emph{AIAA Journal}, pp.\  1--15, 2022.

\bibitem[Zhang(2021)]{zhang2021turbulent}
Zhang, H.
\newblock \emph{Turbulent and Non-Turbulent Interfaces in Low Mach Number Airfoil Flows}.
\newblock PhD thesis, Queen's University (Canada), 2021.

\end{thebibliography}
\bibliographystyle{icml2024}

\end{document}